\documentclass[11pt,aps,superscriptaddress,preprintnumbers,amsmath,amssymb,nofootinbib]{revtex4}
\usepackage{epsfig}  
\usepackage{graphicx}
\usepackage{hyperref}
\usepackage{color}
\usepackage{float}
\usepackage{amsfonts}
\usepackage{amsmath}
\usepackage{slashed}
\usepackage{soul}

\begin{document} 

\begin{flushright}
TIFR/TH/20-39\\
UdeM-GPP-TH-20-284
\end{flushright}

\title{The role of non-universal $Z$ couplings in explaining the $V_{us}$
anomaly}

\author{Ashutosh Kumar Alok}
\email{akalok@iitj.ac.in}
\affiliation{Indian Institute of Technology Jodhpur, Jodhpur 342037, India}

\author{Amol Dighe}
\email{amol@theory.tifr.res.in}
\affiliation{Tata Institute of Fundamental Research, Homi Bhabha Road, \\ Colaba, Mumbai 400005, India}

\author{Shireen Gangal}
\email{shireen.gangal@theory.tifr.res.in}
\affiliation{Tata Institute of Fundamental Research, Homi Bhabha Road, \\ Colaba, Mumbai 400005, India}

\author{Jacky Kumar}
\email{jacky.kumar@umontreal.ca}
\affiliation{Physique des Particules, Universite de Montreal, C.P. 6128, succ.  centre-ville,\\
Montreal, QC, Canada H3C 3J7}

\begin{abstract}
  The tension among  measurements of $V_{us}$ from different
  channels, the so-called Cabibbo Angle Anomaly, can be interpreted
  as a signal of lepton flavor universality (LFU) violation in the
  $W$ boson couplings. 
  We investigate this issue in the framework of effective field theory,
   keeping the  gauge structure of the Standard Model (SM) unchanged.
  We introduce gauge-invariant dimension-6 effective operators that
  couple the Higgs doublet to leptons, thereby giving non-universal
  tree-level contributions to the couplings of electroweak gauge bosons.
  Due to the $SU(2)_L$ gauge symmetry, a tension arises between the $V_{us}$
  measurements that are affected by new $W$ couplings, and the electroweak
  precision measurements, which are also affected by the new $Z$ couplings.
  We show that this tension can be alleviated by allowing additional sources
  of gauge-invariant couplings of $Z$ boson to left- or right-handed leptons,
  and find the optimal regions indicated by the current data in the Wilson-coefficient space.
  We illustrate our model-independent results with the examples of minimal
  extensions of the SM involving the vector-like lepton (VLL) models. 
  We point out that  dimension-6 operators coupling the Higgs doublet to leptons 
  can  affect
  the rate of $h \to \tau \tau$ decay significantly in general, however
  this effect is restricted to less than a per cent level for the minimal VLL
  models. 
\end{abstract}

\maketitle 

 \newpage
 
\section{Introduction}

The  unitarity of the Cabibbo-Kobayashi-Maskawa (CKM) matrix is one of the fundamental
predictions of the Standard Model (SM) of electroweak interactions.
Any deviation from the unitarity of this quark mixing matrix would be an unambiguous
signature of physics beyond the SM.
So far, the CKM paradigm has successfully survived stringent tests in several
precision measurements. However, some tensions have emerged among
determinations of the CKM elements $V_{us}$ and $V_{ud}$ from various sources
\cite{Belfatto:2019swo,Grossman:2019bzp}:

\begin{itemize}
  
\item The element $|V_{us}|$ can be determined from the semileptonic kaon decays
  $K \to \pi \ell \nu$ ($K_{\ell 3}$), where $\ell$ is either an electron or muon, 
  using the vector form factor at zero momentum transfer, $f_+(0)$.
  Using the recent update of $f_+(0)$ from new lattice QCD results with
  $N_f =2+1+1$ flavor \cite{Aoki:2019cca}, $|V_{us}^{K_{\ell 3}}|$ is obtained to
  be $0.22326 \pm 0.00058$ \cite{Antonelli:2010yf,Moulson}.

\item The ratio $|V_{us}/V_{ud}|$  can be determined by comparing $K \to \mu \nu(\gamma)$
  and $\pi \to \mu \nu(\gamma)$ rates and using lattice QCD results for the ratio of
  decay constants, $f_{K}/f_{\pi}$.
  Inclusion of the updated lattice results \cite{Aoki:2019cca} gives
  $|V_{us}/V_{ud}|= 0.23129 \pm 0.00045$ \cite{Antonelli:2010yf,Passemar}.
  Using CKM unitarity, i.e. $|V_{ud}|^2 + |V_{us}|^2 + |V_{ub}|^2 = 1$ (with
  $|V_{ub}|^2 < 10^{-5}$), one gets the value $|V_{us}^{K/\pi}| = 0.22534 \pm 0.00044$.

\item The element $|V_{ud}|$ can be determined from the super-allowed $0^+ \to 0^+$
  nuclear $\beta$ decay predictions, including the short-distance radiative corrections 
  calculated in two schemes: Seng, Gorchtein, Patel, Ramsey-Musolf (SGPR)
  \footnote{The radiative corrections to the neutron $\beta$-decay were re-evaluated using indirect lattice inputs in ref.~\cite{Seng:2020wjq} and were found to be in excellent agreement with the SGPR results.} \cite{Seng:2018yzq,Seng:2018qru}, and Czarnecki, Marciano, Sirlin (CMS) \cite{Czarnecki:2019mwq}.
  The value of $|V_{us}|$ is then obtained using CKM unitarity.
  In the above two schemes, one gets $|V_{ud}|_{\rm SGPR} = 0.97370 \pm 0.00014$ and
  $|V_{ud}|_{\rm CMS} = 0.97389 \pm 0.00018$, respectively. This leads to
  $|V_{ us}^{\beta}|_{\rm SGPR} =0.22782 \pm 0.00062$ and $|V_{us}^{\beta}|_{\rm CMS} =0.22699 \pm 0.00078$. Apart from these electroweak corrections, there can also be nuclear corrections in $0^+ \to 0^+$ transitions \cite{Gorchtein:2018fxl}. These would leave the central values of $|V_{us}^{\beta}|$ unchanged, but would increase the uncertainties. Since the magnitude of this effect is not yet established, we follow  \cite{Crivellin:2020lzu}  where nuclear corrections are not included  in the analysis at this stage, but the possibility of underestimated uncertainties is kept in mind while interpreting the results.

\item The magnitude of $V_{us}$ can also be determined using inclusive and
  exclusive $\tau$ decays.
  The determination of $|V_{us}|$ from inclusive $\tau$ decays to final states involving
  strange quarks is $|V_{us}^{\tau}| = 0.2195 \pm 0.0019$ \cite{Amhis:2019ckw}.
  This extraction of $|V_{us}|$ depends upon corrections due to finite quark masses
  and non-perturbative QCD effects \cite{Gamiz:2004ar,Gamiz:2002nu}. 
  The determination of $|V_{us}^{\tau}|$ from the ratio of decay rates
  $\Gamma(\tau \to K \nu)/\Gamma(\tau \to \pi \nu)$ is $0.2236 \pm 0.0015$,
  while that from $\tau \to K \nu$ decays is  $0.2234 \pm 0.0015 $ \cite{Amhis:2019ckw}.

\end{itemize}

  It is evident that the above measurements are incompatible with each other.
  Compared to the CKM unitarity prediction of $0.2245 \pm 0.0008$
  \cite{pdg}, the $|V_{us}^{\tau}|$ value from the inclusive $\tau$ decays
  is smaller by $\sim 2.9\sigma$, while the average of $|V_{us}^{\tau}| = 0.2221 \pm 0.0013$
  from inclusive and exclusive $\tau$ decays is smaller by $\sim 2\sigma$ \cite{pdg}.
  The $\beta$ decay measurements, on the other hand, yield $|V_{us}^\beta|$ values
  that are higher than the unitarity prediction, the level of inconsistency
  depending upon the radiative corrections scheme. For SGPR and CMS schemes, the combined internal inconsistency of $V_{us}$, $V_{us}/V_{ud}$, and $V_{ud}$ measurements is at the level of 5.1$\sigma$ and 3.6$\sigma$, respectively \cite{Grossman:2019bzp}.

This disagreement,  the ``Cabibbo Angle Anomaly'' (CAA),
can be interpreted as a possible sign for the violation of
CKM unitarity, and has triggered recent interest in potential explanations with
physics beyond the SM \cite{Grossman:2019bzp,Coutinho:2019aiy,Cheung:2020vqm,Crivellin:2020lzu,Endo:2020tkb,Capdevila:2020rrl,Kirk:2020wdk,Crivellin:2020ebi,Crivellin:2020oup,Crivellin:2020klg,Felkl:2021qdn,Belfatto:2021jhf,Branco:2021vhs,Kirk:2021kcs,Chang:2021axw}.  On the other hand, it was argued in \cite{Crivellin:2020lzu} that even while keeping the CKM unitarity intact, this anomaly may be resolved by 
the lepton-flavor universality (LFU) violation in the new couplings
of $W$ bosons to leptons.

In this work, we follow the spirit of \cite{Crivellin:2020lzu}, introducing new $W$-boson couplings to
leptons, while keeping quark couplings unaffected.
We perform a model-independent analysis of non-universal leptonic 
$W$ couplings in the language of Standard Model Effective Field Theory (SMEFT)
that keeps the gauge group structure of SM unchanged.
We restrict ourselves to gauge-invariant dimension-6 operators
$O_{\phi \ell}^{(3)}$, $O_{\phi \ell}^{(1)}$, and $O_{\phi e}$, 
that couple the Higgs doublet to leptons before electroweak
symmetry breaking (EWSB), and in turn give tree-level contributions to the
$W$ and $Z$ couplings after EWSB.
We derive model-independent bounds on these non-universal couplings, in the general
scenario where couplings to all three generations are present.
The bounds are obtained using constraints coming from a number of potentially LFU-violating ratios  in the $B, K, \pi, \tau$ and $\mu$ sectors \footnote{For the sake of brevity, we refer to these ratios as ``LFU ratios''.},
as well as from electroweak precision (EWP) observables. 

Only one of the gauge-invariant dimension-6 operators considered here,
$O_{\phi \ell}^{(3)}$, contributes to new $W$ couplings that are essential
for resolving the CAA. 
However since this operator also contributes to leptonic couplings of
the $Z$ boson, the EWP measurements severely restrict a deviation from the SM
if $O_{\phi \ell}^{(3)}$ is the only operator, and we find that the tension between
the solution for CAA and the constraints from EWP persists.
We perform a systematic study of the reduction
of this tension by the addition of the other operators $O_{\phi \ell}^{(1)}$
and $O_{\phi e}$, which give rise to new additional couplings of the $Z$
boson to left-handed and right-handed leptons, respectively.
We find the optimal conditions for the ratios of Wilson coefficients (WCs) of these
operators needed to resolve the tension, in a model-independent analysis.
The favoured values of the ratios of WCs, obtained from this
analysis, would act as a guide for the construction of models.

Minimal extensions of the SM that add only one species of new vector-like
leptons (VLL) to the SM particle content are prime examples of the
models that give rise to strongly correlated $O_{\phi \ell}^{(1)}$
and $O_{\phi \ell}^{(3)}$ operators.  
In addition to the LFU and EWP constraints considered for the model-independent
analysis, these models also get constrained from lepton-flavor violating (LFV)
processes $\ell_i \to \ell_j \bar\ell_k \ell_k$.
In the context of minimal VLL models $N$ and $E$ that involve $SU(2)_L$ singlets,
and models $\Sigma$ and $X$ that involve $SU(2)_L$ triplets, 
we illustrate that our conclusions from the model-independent analysis stay
valid -- the closer the ratio of WCs in a model is to the
optimal value predicted by the model-independent
analysis, the better is the model in reducing the tension between $V_{us}$ and
EWP observables, and hence in resolving the CAA.

The tension between $V_{us}$ measurements  and EWP data in the presence of a single-operator dominance was pointed out in \cite{Kirk:2020wdk}. In this work, we have gone a step ahead and shown how this tension may be resolved in a model-independent fashion. In ref.~\cite{Crivellin:2020ebi}, the CAA anomaly was analyzed  in the context of the VLL models.    However, our approach is complementary -- we find the optimal ratios of WCs of new operators in a model-independent manner, and use the
  minimal VLL models as examples to  validate our model-independent results.

The operator $O_{\phi \ell}^{(3)}$ would affect the measurement of the Fermi constant $G_F$, and hence the inferred charged-lepton Yukawa couplings. This would spoil the SM relationship between the charged-lepton Yukawa couplings and the decay rate of the Higgs boson to these charged leptons. The extent of this effect is severely restricted by measurements involving intermediate $W$ and $Z$ bosons.
However, there exists a dimension-6 operator $O_{\phi \ell e}$
that contributes neither to the $W$ couplings nor to the $Z$ couplings, but
can influence the couplings of leptons with the Higgs boson. We explore
the effect of this operator on $h \to \tau \tau$ decay. We show that, while significant deviations of this decay rate from the SM prediction is possible in model-independent schemes,  the minimal VLL models considered above cannot change this decay rate by more than a per cent, and hence any significant ($> 1\%$) deviation of these measurements from
the SM would confirm the need to go  beyond the minimal VLL framework.

Our work is organized as follows: in Sec.~\ref{formalism}, we introduce the new
gauge-invariant dimension-6 operators that would give rise to tree-level
$W$ and $Z$ couplings after EWSB. 
In Sec.~\ref{Constraints}, we discuss the model-independent constraints from
$V_{us}$ measurements, LFU ratios, EWP observables, and LFV decays.
In Sec.~\ref{results-eft}, we show our model-independent results when only
$O_{\phi \ell}^{(3)}$ is present, as well as when it is present in combination
with $O_{\phi \ell}^{(1)}$ or $O_{\phi e}$. In the scenarios where two operators are
present simultaneously, we find the optimal ratios of WCs of these operators that would resolve the CAA anomaly without conflicting with the EWP data.
In Sec.~\ref{sec:miminal-ext-sm}, we
exemplify the model-independent results with two of the minimal VLL models.
The analysis of the leptonic Higgs boson decays in the presence of the
operator $O_{\phi \ell e}$ is performed in Sec.~\ref{sec:higgs}.
Finally, we conclude in Sec.~\ref{sec:conc}.

\section{Model-independent formalism}
\label{formalism}

In the SMEFT, the Standard Model is extended by higher dimensional operators
$O_a$ that are  $SU(3)_C \times SU(2)_L \times U(1)_Y$  gauge invariant.
The SMEFT Lagrangian can be expressed as
\begin{align}
  \mathcal{L}_{\rm SMEFT}^{d \ge 5} = \sum_{O_a^\dagger = O_a} C_a O_a + \sum_{O_a^\dagger \ne O_a}
  \left ( C_a O_a + C_a^* O_a^\dagger  \right   )\,,
  \label{eq:Lagrangian}
\end{align}
where the dimensionful coefficients $C_a$ are known as WCs.
We restrict ourselves to dimension-6 operators which modify the couplings of
charged leptons and neutrinos to $W$ and $Z$ bosons after EWSB
\cite{Buchmuller:1985jz,Grzadkowski:2010es}. There are three such operators that
arise from the couplings of leptons to the Higgs doublet:
\begin{align}
  [O_{\phi \ell}^{(1)}]_{ij} &=
  (\phi^\dagger i \,{\overleftrightarrow {D_\mu}} \, \phi)(\bar{\ell}_i \, \gamma^\mu\, \ell_j)\,,
  \label{eq:O1}\\ 
  [O_{\phi \ell}^{(3)}]_{ij} &=
  (\phi^\dagger i\, {\overleftrightarrow {D_\mu^I}} \, \phi) (\bar \ell_i \tau^I \gamma^\mu \ell_j),
  \label{eq:O3} \\
  [O_{\phi e}^{}]_{ij} & =
  (\phi^\dagger i\, {\overleftrightarrow {D_\mu}}\, \phi)(\bar{e}_i \, \gamma^\mu\, e_j)\,,
  \label{eq:Ophie}
\end{align}
where $\phi$ is the Higgs doublet, $\ell$ is the left-handed lepton doublet,
and $e$ is the right-handed lepton singlet, under $SU(2)_L$. The indices $i$ and $j$ correspond to lepton generations.
The covariant derivative is
$D_\mu = \partial_\mu + i g_2 W_\mu^a \dfrac{\tau^a}{2} + i g_1 B_\mu Y$.

The operator $C_{\phi \ell}^{(3)}$ gives corrections to the couplings of $W$ and $Z$ bosons
to the left-handed leptons, whereas the operators $C_{\phi \ell}^{(1)}$ and $C_{\phi e}$
modify the $Z$-couplings to left-handed and right-handed leptons, respectively.  
We can parameterize the charged-current (CC) new physics (NP) contribution as 
\begin{equation}
	\label{ew:W-couplings}
\begin{aligned}
\delta \mathcal{L}_{W}^{\rm NP} &= 
	-\frac{g_2}{\sqrt{2}} \varepsilon_{ij} \bar{\ell}_i \gamma^\mu P_L  \nu_j W_\mu^- ~+ H.c.\,,
\end{aligned}
\end{equation}
where the dimensionless parameter $\varepsilon_{ij}$ is given by \cite{Crivellin:2020ebi}
\begin{equation}
	\label{eq:eps}
	 \varepsilon_{ij}  = v^2 ~ [C_{\phi \ell}^{(3)}]_{ij} \; .
\end{equation}
Here $v = 246$ GeV is the vacuum expectation value of the Higgs field.
It is worth noting that the diagonal elements of $\varepsilon_{ij}$ are real, whereas the 
off-diagonal elements can take complex values in general. 
Similarly, the $Z$-couplings in terms of WCs can be parameterized as
\begin{equation}
\begin{aligned}	
  \delta  \mathcal{L}_{ Z}^{\rm NP} & = -\frac{g_Z}{{2}} Z_{\mu}  \Bigg[
  \bar \ell_i \, \gamma^{\mu} \left(
        [g_L^{\ell}(Z)]_{ij} \, P_L
        \,+\, [g_R^{\ell}(Z)]_{ij} \, P_R \right) \ell_j \\
       & + \bar \nu_i \, \gamma^{\mu} [g_L^{\nu}(Z)]_{ij} \, P_L  \nu_j \Bigg] \; .
\end{aligned}
\end{equation}
Here  
\begin{equation}
  \label{eq:Z-couplings}
\begin{aligned}
  \phantom{x}[g^\nu_L(Z)]_{ij} &
	=  v^2 \left[C_{\phi \ell}^{(1)} - C_{\phi \ell}^{(3)} \right ]_{ij}   \,,\\
  [g^\ell_L(Z)]_{ij} &
	=  v^2 \left[C_{\phi \ell}^{(1)} + C_{\phi \ell}^{(3)}\right]_{ij} \,, \\
  [g^\ell_R(Z)]_{ij} &
	=  v^2 [C_{\phi e}]_{ij}\,,
\end{aligned}
\end{equation}
where  $g_Z=\sqrt{g_1^2+g_2^2}$.

It is important to note here that, due to the $ SU(2)_L$ gauge invariance,
the operator $C_{\phi \ell}^{(3)}$ giving NP contribution to the leptonic
$W$-boson couplings in Eq.~\eqref{eq:eps} also contributes to the
$Z$-boson couplings to left-handed leptons as given in Eq.~\eqref{eq:Z-couplings}. Therefore, 
in the presence of this operator, the extraction of $V_{us}$ from the charge current
decays is inevitably connected to the EWP observables that constrain the properties
of the $Z$ boson. However, since $C_{\phi \ell}^{(1)}$ and $C_{\phi e}$
affect only the $Z$ couplings, this correlation can be broken if NP generates these 
operators as well.
On the other hand, in the context of specific models, the WCs of
above operators may be related, and the correlation between $W$ and $Z$ couplings
may not be broken, but just changed in a predictable way.

\section{Constraints}
\label{Constraints}

In this section, we consider the relevant observables which provide constraints on the 
modified $W$ and $Z$ couplings. 
For this we include the $|V_{us}|$ measurements, several LFU ratios, and
EWP observables, which constrain the diagonal elements of the matrices of
WCs.
We also include LFV observables that constrain the off-diagonal WCs.

\subsection{$|V_{us}|$ constraints}
\begin{table}[h!]
  \begin{center}
\begin{tabular}{c|c}
\toprule
Observable  & Experimental Value \\ \hline
 $R(V_{us})$ &   $0.9891 \pm 0.0033$ \cite{Crivellin:2020lzu} \\
  $|V_{us}^{\tau}|$ &  $0.2221 \pm 0.0013$  \cite{Amhis:2019ckw} \\
\hline
\end{tabular}
\caption{Experimental values of $|V_{us}|$ observables.}
 \label{vus-constraint}
  \end{center}
\end{table}

The determination of $|V_{us}|$, as discussed briefly in the introduction, mainly
comes from three sources: kaon decays, super-allowed beta decays, and tau decays.
The ratio of the branching fractions of purely muonic kaon decay ($K \to \mu \nu$)
and pion decay ($\pi \to \mu \nu$) is used to determine $|V_{us}|/|V_{ud}|$.
However this quantity, being a ratio of $|V_{us}|$ and $|V_{ud}|$, is independent of
the anomalous $W\ell\nu$ couplings, and hence of the $\varepsilon_{ij}$ parameter.
The value of $|V_{us}|$ extracted from this ratio is close to the PDG average. 
The modified $W\ell\nu$ couplings, however, affect the Fermi constant extracted from
the muon decay process $\mu \to e \nu_e \nu_\mu$ as \cite{Crivellin:2020lzu}
\begin{align}
  G_F = G_F^\mathcal{L} ~\Big(1 + \varepsilon_{ee}
  + \varepsilon_{\mu \mu} \Big) \; ,
\end{align}
where $G_F^\mathcal{L} \equiv (\sqrt{2} v^2)^{-1}$ is the Fermi constant
in the SM. Since $G_F$ enters most of the EWP observables, the couplings
$\varepsilon_{ii}$ are strongly constrained. 
The determination of $|V_{us}|$ from semi-leptonic kaon decay $K \to \pi \mu \nu$
is sensitive to LFU-violating couplings, through the modification of 
the Fermi constant $G_F$ and the anomalous $W$ couplings \cite{Crivellin:2020lzu}:
\begin{align}
|V_{us}^{K_{\mu_3}}| = |V_{us}^\mathcal{L}| \left( 1 - \varepsilon_{ee} \right) \;  ,
\end{align}
where $|V_{us}^\mathcal{L}|$ denotes the CKM matrix element in the SM.

The determination of $|V_{ud}|$ from $\beta$ decays is also affected due to the
redefinition of $G_F$ as \cite{Crivellin:2020lzu}
\begin{align}
  |V_{us}^\beta| = \sqrt{1- |V_{ud}^\beta|^2 - |V_{ub}|^2} \,
  \approx \, |V_{us}^\mathcal{L}|~
  \Big[1 + \Big|\frac{V_{ud}^\mathcal{L}}{V_{us}^\mathcal{L}}\Big| \,\varepsilon_{\mu \mu} \Big]\,.
\label{eq:Vusbeta}
\end{align}
The observable $R(V_{us})$ as defined in \cite{Crivellin:2020lzu} is sensitive to LFU
violation, as it is a ratio of $|V_{us}|$ measured from kaon decays involving only muons
to that determined from beta decays which involve only electrons.
Using Eq.~(\ref{eq:Vusbeta}), this quantity is
given by
\begin{align}
  R(V_{us}) = \Big|\frac{V_{us}^{K_{\mu_2}}}{V_{us}^\beta}\Big| \, \approx \,
  1- \Big|\frac{V_{ud}}{V_{us}}\Big|^2\, \varepsilon_{\mu\mu}
  \, \approx  \, 1- 20 ~ \varepsilon_{\mu \mu} \,.
\end{align}
Note that the large sensitivity to $\varepsilon_{\mu \mu}$ is the effect of the
hierarchy between $V_{us}$ and $V_{ud}$ magnitudes. Here, we use the SGPR scheme \cite{Seng:2018qru} for calculating radiative corrections in beta decay calculations.

The value of $|V_{us}^\tau|$ determined from inclusive tau decays and from the ratio
$\Gamma(\tau\to K \nu)/ \Gamma(\tau \to \pi \nu)$ is insensitive to the modified
$W\ell\nu$ couplings.  However,
its determination from the exclusive decay $\tau \to K \nu$ depends on all three
$\varepsilon_{ii}$ parameters, and is given by
\begin{align}
  |V_{us}^{\tau}| = |V_{us}^\mathcal{L}| ~ \left(1 - \varepsilon_{ee}
  -\varepsilon_{\mu \mu} + \varepsilon_{\tau \tau} \right)\,.
\end{align}

In our fit, we include the observable $R(V_{us})$, and $|V_{us}^\tau|$ from $\tau \to K \nu$
decays, to examine the consistency of these measurements with the constraints from
EWP and LFU decays in the parameter space of the WCs
$C_{\phi\ell}^{(3)}, \, C_{\phi\ell}^{(1)},\, C_{\phi e}$. The experimental values of these observables are listed in Table~\ref{vus-constraint}.

\subsection{EWP constraints}

\begin{table*}[h!]
  \begin{center}
\begin{tabular}{cc|cc}
\toprule
Observable &  Experimental value & Observable & Experimental value \\ \hline
	$\sigma_{h}$& $ 41.4742 \pm 0.0326 $ (nb) \cite{ALEPH:2005ab} & $A_{FB}$($Z\to bb$) & $ 0.0992 \pm 0.0016$ \cite{ALEPH:2005ab}\\ 
	$R_e$& $20.8038 \pm 0.0497$ \cite{ALEPH:2005ab} & $A_{FB}$($Z\to cc$)& $0.0707 \pm 0.0035$ \cite{ALEPH:2005ab} \\ 
$R_\mu$& $20.7842 \pm 0.0335$ \cite{ALEPH:2005ab} & A($Z\to bb$)&  $0.923 \pm 0.020$ \cite{ALEPH:2005ab}\\ 
$R_{\tau}$& $20.7644 \pm 0.0448$ \cite{ALEPH:2005ab} & A($Z\to cc$)& $0.670 \pm 0.027$ \cite{ALEPH:2005ab} \\ 
	$A_{FB}$($Z\to ee$)& $0.0145 \pm 0.0025$ \cite{ALEPH:2005ab} & $\Gamma_Z$& $2.4955 \pm 0.0023$ \cite{ALEPH:2005ab} \\ 
	$A_{FB}$($Z\to \mu \mu$)&  $0.0169 \pm 0.0013$ \cite{ALEPH:2005ab} & $m_W$& $  80.387 \pm 0.016$ \cite{Aaboud:2017svj,Aaltonen:2013iut}\\ 
	$A_{FB}$($Z\to \tau \tau$)& $0.0188 \pm 0.0017$ \cite{ALEPH:2005ab} & $\Gamma_W$& $2.085 \pm 0.042 $ \cite{pdg} \\ 
A($Z\to ee$)& $  0.1513 \pm 0.0019 $ \cite{ALEPH:2005ab} &  BR($W\to e\bar \nu$)& $ 0.1071 \pm 0.0016$ \cite{Schael:2013ita}\\ 
A($Z\to \mu\mu$)&  $0.142 \pm 0.015$ \cite{ALEPH:2005ab} & BR($W\to \mu\bar \nu$)& $0.1063 \pm 0.0015$ \cite{Schael:2013ita} \\ 
A($Z\to \tau \tau)$& $0.1433 \pm 0.0043$ \cite{ALEPH:2005ab} & BR($W\to \tau\bar \nu$)& $0.1138 \pm 0.0021$ \cite{Schael:2013ita} \\ 
$R_b$& $ 0.21629 \pm 0.00066$ \cite{ALEPH:2005ab} & BR($W\to cX$)& $0.49 \pm 0.04$ \cite{pdg} \\
$R_c$& $  0.1721 \pm 0.0030$ \cite{ALEPH:2005ab} &$R_{\mu e}(W\to \ell \bar \nu)$& $0.980 \pm 0.002 \pm 0.018$ \cite{Aaij:2016qqz} \\ 
$A(Z\to ss)$&  $ 0.895 \pm 0.066 \pm 0.062$ \cite{Abe:2000uc} & $R_{\tau e}(W\to \ell \bar \nu)$&  $ 0.960 \pm 0.062$ \cite{Abbott:1999pk} \\ 
	&  & $R_{\tau \mu}(W\to \ell \bar \nu)$ &  $ 0.992 \pm 0.013$ \cite{Aad:2020ayz} \\ \hline
\end{tabular}
\caption{List of EWP observables and their experimental values used in this analysis.}
\label{table-ew}
\end{center}
\end{table*}

The operator $O_{\phi \ell}^{(3)}$ in Eq.~\eqref{eq:O3}, required to introduce
new $W$-couplings ($\varepsilon_{ij}$) in a gauge invariant way, also
modifies the left-handed $Z$-boson couplings. 
Therefore, one has to consider various EWP observables 
which have been measured with high precision at LEP. 
These observables depend on three main parameters: the fine structure constant $\alpha$,
the mass of the $Z$ boson $(M_Z)$, and the Fermi constant $(G_F)$. 
We closely follow the strategy used in Ref. 
\cite{Aebischer:2018iyb}, where 27 observables are considered in the EW precision 
global fit. For completeness, we list these observables in Table~\ref{table-ew}.

 \subsection{LFU constraints}
 
 The LFU ratios provide constraints on the diagonal elements $\varepsilon_{ii}$. 
 If these diagonal elements are different from each other, then LFU violation is
 present. Such a violation can be tested by defining ratios of branching fraction
 of decays involving different leptons in the final state. 
 The off-diagonal elements can also affect the LFU ratios, however their effect appears only at the second order in $\varepsilon$, and hence
 they are suppressed.  
 The LFU observables used in our analysis and their dependence on the $\varepsilon_{ij}$
 parameters is listed in Table~\ref{table-lfu}.
 
 
 \begin{table*}[h]
   \begin{center}
\begin{tabular}{c|c|c}
\toprule
	\quad\quad LFU Ratio \quad\quad & Experimental value &  Dependence on $\varepsilon_{ij}$    \\
 \hline
	$R_{K}^{\mu e}\equiv |\mathcal{A}(K\rightarrow\mu\nu)|/ |\mathcal{A}(K\rightarrow e\nu)|$ \quad \quad & $0.9978 \pm 0.0018$\cite{Pich:2013lsa} &  \\
	$ R_{K\pi}^{\mu e}\equiv |\mathcal{A}(K\rightarrow\pi\mu\bar{\nu})|/ |\mathcal{A}(K\rightarrow \pi e\bar{\nu})|$ \quad \quad & $1.0010 \pm 0.0025$ \cite{Pich:2013lsa}  &  $ 1+  {\rm } \varepsilon_{\mu\mu}-{\rm }\varepsilon_{ee} $  \\
	$R_{B D}^{\mu e} \equiv |\mathcal{A}(B\rightarrow D^{(*)}\mu\nu)|/ |\mathcal{A}(B\rightarrow D^{(*)}e\nu)|$ \quad  \quad& $0.9890 \pm 0.0120 $\cite{Jung:2018lfu}  &  \\
	$ R_{\pi}^{\mu e}\equiv |\mathcal{A}(\pi\rightarrow\mu\nu)|/ |\mathcal{A}(\pi\rightarrow e\nu)|$ \quad \quad&   $1.0010 \pm 0.0009$\cite{Amhis:2019ckw}&  \\
	$ R_{\tau \tau}^{\mu e} \equiv |\mathcal{A}(\tau\rightarrow\mu\nu\bar{\nu})|/ |\mathcal{A}(\tau\rightarrow e\nu\bar{\nu})| $  \quad \quad& $1.0018 \pm 0.0014$ \cite{Amhis:2019ckw, Tanabashi:2018oca} &  \\
\hline
	$ R_{\tau \mu}^{e e}\equiv |\mathcal{A}(\tau\rightarrow e\nu\bar{\nu})|/ |\mathcal{A}(\mu\rightarrow e\bar{\nu}\nu)| $ \quad \quad & $ 1.0010 \pm 0.0014$  \cite{Amhis:2019ckw, Tanabashi:2018oca}&  \\
	$R_{\tau \pi}^{\pi \mu }\equiv |\mathcal{A}(\tau\rightarrow \pi\nu)|/ |\mathcal{A}(\pi\rightarrow \mu\bar{\nu})|$ \quad \quad& $ 0.9961 \pm 0.0027$\cite{Amhis:2019ckw}   &  $1+{\rm}\varepsilon_{\tau\tau}-{\rm }\varepsilon_{\mu \mu}$  \\
	$R_{\tau K}^{K \mu }\equiv |\mathcal{A}(\tau\rightarrow K\nu)|/ |\mathcal{A}(K\rightarrow \mu\bar{\nu})|$ \quad \quad& $0.9860  \pm 0.0070$ \cite{Amhis:2019ckw} &  \\
\hline
	 $R_{\tau \mu}^{ \mu e }\equiv |\mathcal{A}(\tau\rightarrow \mu\nu\bar{\nu})|/ |\mathcal{A} (\mu\rightarrow e\nu\bar{\nu})|$ \quad \quad &$1.0029  \pm 0.0014$ \cite{Amhis:2019ckw, Tanabashi:2018oca} &  $1+{\rm}\varepsilon_{\tau\tau}-{\rm }\varepsilon_{ee}$ \\
\hline
\end{tabular}
\caption{LFU ratios in the $K$, $B$ , $\pi$, $\mu$ and $\tau$ decays and their dependence on the new effective leptonic $W$ couplings $\varepsilon_{ii}$.
}. 
 \label{table-lfu}
  \end{center}
\end{table*}

\subsection{LFV constraints}
  
The charged LFV decays are induced at the tree level in the EFT
due to the off-diagonal elements of the modified $W$ and $Z$ couplings, $\varepsilon_{ij}$. 
We consider the LFV decays $\ell_i \to \ell_j \bar\ell_k \ell_k$,  for which the current experimental upper
bounds \cite{Bellgardt:1987du,Hayasaka:2010np} are as given in Table~\ref{table-lfv}.

\begin{table*}[h]
  \begin{center}
\begin{tabular}{c|c}
\toprule
LFV Decay & Upper Bound \\ \hline
${\rm Br}\!\left(\mu^+\rightarrow e^+ e^- e^+ \right)$ &   $1.0 \times 10^{-12}$ \\
${\rm Br}\!\left(\tau^- \rightarrow e^- e^+ e^- \right)$  &  $2.7\times 10^{-8}$  \\
${\rm Br}\!\left(\tau^- \rightarrow \mu^- \mu^+ \mu^- \right)$ &  $2.1\times 10^{-8} $ \\
${\rm Br}\!\left(\tau^- \rightarrow e^- \mu^+ \mu^- \right)$ &  $2.7 \times10^{-8}$  \\
${\rm Br}\!\left(\tau^- \rightarrow \mu^- e^+ e^- \right)$ &  $1.8\times 10^{-8}$  \\
${\rm Br}\!\left(\tau^- \rightarrow \mu^- e^+ \mu^-  \right)$ & $1.7\times 10^{-8}$  \\
${\rm Br}\!\left(\tau^- \rightarrow e^- \mu^+ e^- \right)$  & $1.5\times 10^{-8}$ \\
\hline
\end{tabular}
\caption{LFV decays and their experimental upper bounds
   at 90\% C. L. \cite{Bellgardt:1987du,Hayasaka:2010np}.}
 \label{table-lfv}
  \end{center}
\end{table*}
 
Using these, we obtain the following limits at
95\% C.L.: 
\begin{align}
\begin{split} 
|\varepsilon_{e\mu}|\leq 0.3\times10^{-5},\,\,\,\,
|\varepsilon_{e \tau }|\leq 0.9\times10^{-3},\,\,\,\,
|\varepsilon_{\mu\tau}|\leq 0.9\times10^{-3}.
\end{split}
\end{align} 
These limits do not affect our model-independent analysis,
however they can be used to put constraints on the off-diagonal couplings
in the VLL models that we consider later in our analysis.

\section{Preferred NP couplings in the model-independent formalism}
\label{results-eft}

In order to examine
the level of consistency among different observables, and to identify the allowed parameter space for the WCs, we perform a
$\chi^2$ analysis. The function $\chi^2(\{C_i\})$, where $\{C_i\}$ is the
set of all relevant WCs, is constructed as
\begin{equation}
  \chi^2 (\{C_i\}) =\sum
  \left[ {\cal O}_{\rm th}(\{C_i\}) - {\cal O}_{\rm exp} \right]^T {\cal C}^{-1}
  \left[ {\cal O}_{\rm th}(\{C_i\}) - {\cal O}_{\rm exp} \right] \; ,
\end{equation}  
where the sum is over the measurements relevant for $R(V_{us})$, EWP observables,
and LFU ratios.
Here ${\cal O}_{\rm th}(\{C_i\})$ are the theoretical predictions of the observables
at the given values of $\{C_i\}$,
while ${\cal O}_{\rm exp}$ are experimental measurements. 
The covariance matrix ${\cal C}$ is obtained by adding the individual theoretical
and experimental covariance matrices that take care of the correlations
among different observables. 
The theoretical predictions are obtained using publically available packages
{\tt flavio} \cite{Straub:2018kue} along with {\tt wilson} \cite{Aebischer:2018bkb}.
We have implemented additional observables such as $R(V_{us})$ and LFU ratios
in {\tt flavio}, since these are not available in {\tt flavio}.
The minimization of the $\chi^2(\{C_i\})$ function is performed using the CERN
library {\tt MINUIT} \cite{James:1975dr}.
 We find $\chi^2_{\rm SM}=50.77$ for the SM value, i.e. for vanishing WCs.
This corresponds to 38 degrees of freedom.

As discussed in section \ref{formalism}, there are in total three types of
SMEFT WCs  which give tree-level contribution to the $W$ and $Z$ couplings:
$C_{\phi \ell}^{(3)}$,  $C_{\phi \ell}^{(1)}$, and $C_{\phi e}$.
Out of these, non-zero $C_{\phi \ell}^{(3)}$ are essential if we want to
address the CAA, since it is the only WC that would contribute to the
NP leptonic CC couplings. 
We allow NP couplings to all three flavors of leptons.
We consider two classes of EFT scenarios:

\begin{itemize}

\item Minimal EFT scenario: \\
  In this case, we only allow a single SMEFT operator, $O_{\phi \ell}^{(3)}$, which 
  affects the $W$-couplings to left-handed leptons, parameterized by
  $\varepsilon_{ij}$ as in Eq.~\eqref{eq:eps}. 
  However, this operator also gives rise to corrections to the $Z$-boson couplings to
  left-handed leptons as shown in Eq.~\eqref{eq:Z-couplings}, and these NP leptonic $Z$ couplings
  are highly correlated to the NP leptonic $W$ couplings.

\item Non-minimal EFT scenarios: \\
  In order to break the strong correlations between $W$ and $Z$ couplings present in
  the minimal scenario,  we allow NP in the form of  additional operators $C_{\phi \ell}^{(1)}$ or
  $C_{\phi e}$ which give tree level contributions to the  $Z$ boson couplings with
  left- and right-handed leptons, respectively, through Eq. \eqref{eq:Z-couplings}.
  Specifically, we focus on two simple cases satisfying the following relations:
 \begin{align}\label{eq:non-minimal-eft}
   {\rm I:} \quad C_{\phi \ell}^{(1)} &= \alpha ~C_{\phi \ell}^{(3)}\,, \\
   {\rm II:} \quad C_{\phi e} &= \beta ~ C_{\phi \ell}^{(3)}\,,
\label{eq:non-minimal-eft2} \end{align}
 where $\alpha$ and $\beta$ are free parameters which control the size of couplings
 of $Z$ to left- and right-handed leptons, respectively, relative to the $W$ boson
 couplings to left-handed leptons. Note that there is still quite a strong correlation among the WCs, since the ratios of WCs, $\alpha$ and $\beta$, are taken to be flavor-independent. However, such a relation is motivated from many models, especially from the VLL models that will be considered in the following sections.
\end{itemize}

\subsection{Minimal EFT scenario}

%
\begin{figure*}[hbt!]
       \begin{center}
\includegraphics[width=0.45\textwidth]{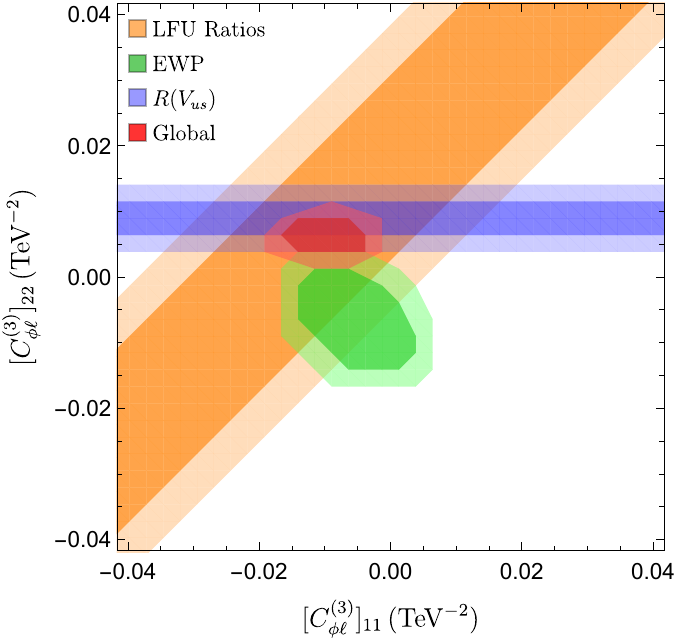}
\hspace{0.65cm}
\includegraphics[width=0.45\textwidth]{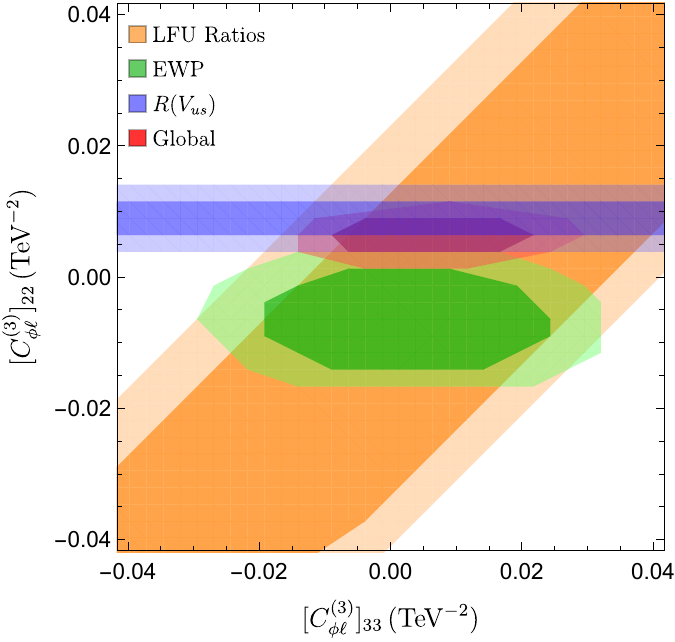} 
\\ 
\vspace{0.5cm}

\includegraphics[width=0.47\textwidth]{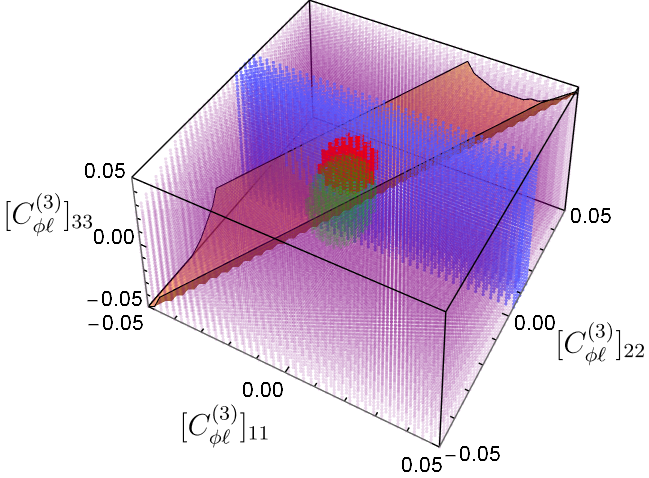}
\hspace{0.5cm}
\includegraphics[width=0.45\textwidth]{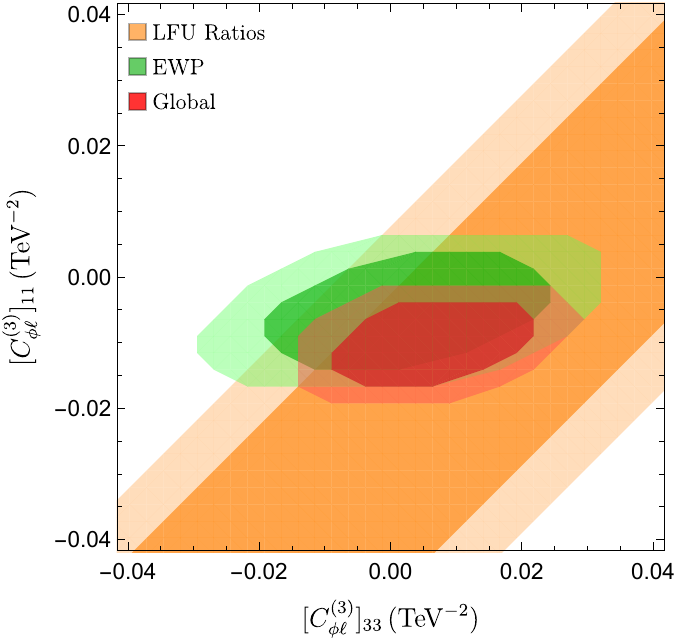}
	       \caption{Model-independent constraints in the parameter spaces
  of the WCs $[C_{\phi \ell }^{(3)}]_{ii}$ at $68\%$ and $95\%$ C.L.
  for two-parameter fits ($\Delta \chi^2 =$ 2.3 and 4.6, respectively).
  The top-left, top-right, and bottom-right panels show projections onto
  three 2D planes (referred to in the text as $e-\mu$, $\tau-\mu$, and $\tau-e$
  planes, respectively), while the bottom-left panel shows a 3D view.
  Note that the $R(V_{us})$ constraint, visible as a vertical blue slab in the 3D
  view, allows the whole parameter space in the $\tau-e$ plane, and hence has not been explicitly shown. Similarly, the constraint from $|V_{us}|$ measurement from exclusive $\tau$ decays is
  visible only in the 3D view, as an inclined purple slab.
}
\label{fig:mi}
       \end{center}
\end{figure*}

The allowed parameter space for the WCs $[C_{\phi \ell}^{(3)}]_{ii}$ for $i=1,2,3$
(corresponding to $e,\mu,\tau$ flavors, respectively), obtained by using the
$R(V_{us})$, EWP and LFU constraints discussed in the previous section,
are shown in Fig.~\ref{fig:mi}.
While depicting the 2D projections for the three orthogonal views, 
minimization of $\chi^2$ over the third direction is performed.
The dark and light colors correspond to
$\Delta \chi^2_{\rm} \equiv \chi^2 -\chi^2_0$ =  2.3 and 4.6, respectively,
where $\chi^2_0$ corresponds to the best-fit point.
These projections thus correspond to the 68\% and 95\% C.L. intervals
for these two-parameter constraints.

It may be seen from the figure that the EWP data strongly constrain
the NP parameter space, keeping it near the SM point, i.e. near 
$[C_{\phi \ell}^{(3)}]_{11} = [C_{\phi \ell}^{(3)}]_{22} =[C_{\phi \ell}^{(3)}]_{33} =0$.
The other constraints, on the other hand, allow a sizeable deviation from the SM. 
The favoured regions due to the LFU measurements are sensitive to the differences
$\varepsilon_{ii} - \varepsilon_{jj}$, and hence lie along a diagonal in the
$e-\mu$, $\tau-\mu$ and $\tau-e$ planes.
Since the determination of $|V_{us}|$ from $\tau$ decays is controlled by the
combination $(-\varepsilon_{11} - \varepsilon_{22} +  \varepsilon_{33})$,
the region favoured by this measurement is an inclined plane in 
the 3D parameter space.
The $R(V_{us})$ measurement is the one that demands non-zero NP couplings, 
$[C_{\phi \ell}^{(3)}]_{22} >0$.
The net global best fit is at
 $([C_{\phi \ell}^{(3)}]_{11}, [C_{\phi \ell}^{(3)}]_{22} , [C_{\phi \ell}^{(3)}]_{33}) =
  (-9.5, 6.3, 6.7)\times 10^{-3}$, whereas the SM is disfavoured at
 $\Delta \chi^2 \approx  11.0$. While negative sign for $[C_{\phi \ell}^{(3)}]_{11}$ and positive sign for $[C_{\phi \ell}^{(3)}]_{22}$ are preferred in the global fit, large errors in $[C_{\phi \ell}^{(3)}]_{33}$ allow it to have either sign.

Note that, though the fit seems to have improved with the introduction of
NP parameters, there is still a  clear tension between the EWP and $R(V_{us})$ 
measurements. The $2\sigma$-favoured regions for these two sets of
measurements barely overlap in the 3D NP parameter space. 
On the other hand, these two sets of measurements are individually
compatible with the constraints from the LFU ratios and $\tau$ decays within $1\sigma$.
The region allowed by a combination of LFU ratios and EWP data favors negative
values of $[C_{\phi \ell}^{(3)}]_{22}$, while the region preferred by $R(V_{us})$
and LFU ratios combined favors positive values of this parameter.
This tension between the EWP and $R(V_{us})$ measurements is the reason
why the improvement offered by this scenario over the SM is only marginal.

The reason behind the failure of this minimal scenario to resolve the CAA
lies in the strong correlation between the values of the effective $W$ and $Z$ couplings 
(see eqs.~\eqref{eq:eps} and \eqref{eq:Z-couplings})
 in the minimal scenario.  
Clearly, different regions of parameter space are preferred by the $W$ couplings
required to accommodate $|V_{us}|$, i.e. $[C_{\phi \ell}^{(3)}]_{22} >0$, and by the 
$Z$ couplings required to be consistent with the $Z$-pole observables:
$[C_{\phi \ell}^{(3)}]_{22} \leq 0$. 

It is therefore evident that the single operator $O_{\phi l}^{(3)}$ by itself cannot 
account for the present $|V_{us}|$ data while satisfying the EWP measurements. 
Hence, the presence of additional NP operators is desirable. Since the other
NP operators $C_{\phi \ell}^{(1)}$ and $C_{\phi e}$ can influence the $Z$ couplings
without affecting the $W$ couplings, they  break the strong
correlation between these couplings, thus allowing us to get a better fit.
The presence of these additional operators is actually quite natural, since
the symmetries of SMEFT allow all these operators to be present, and one would
have needed a special reason for the absence of any of these operators.

\subsection{Non-minimal EFT scenarios}
\label{sec:non-minimal}

\begin{figure*}[h]
       \begin{center}
\includegraphics[width=0.45\textwidth]{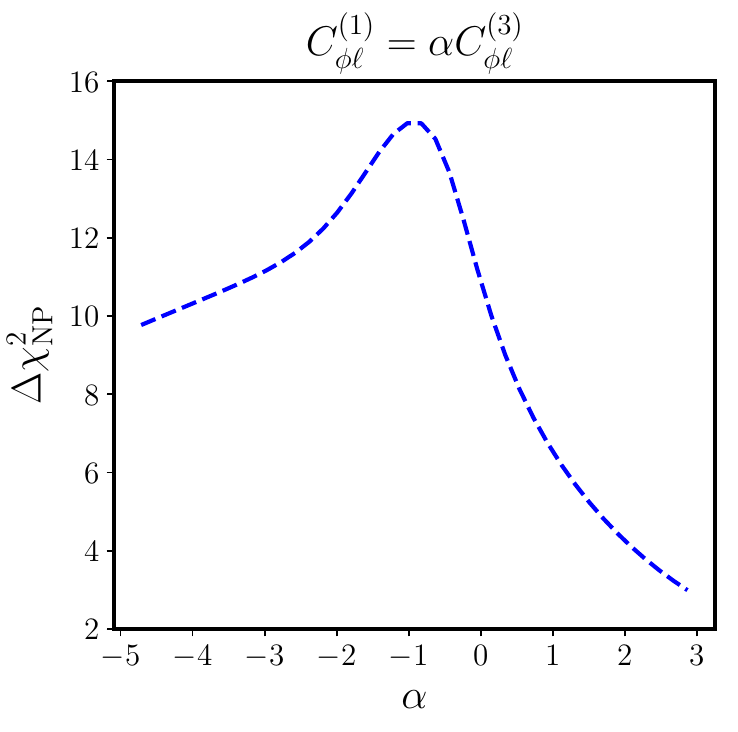}
\hspace{0.5cm}
\includegraphics[width=0.45\textwidth]{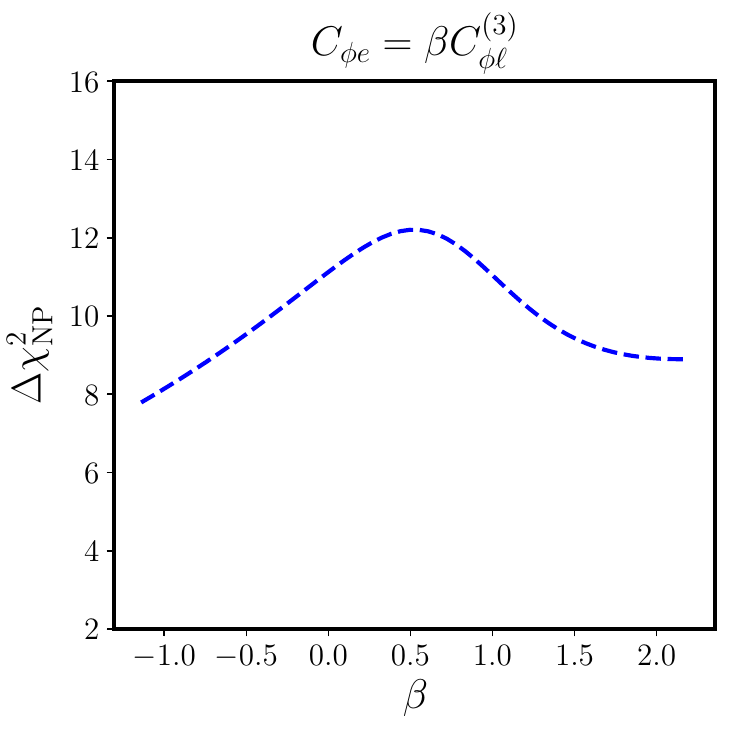}
\caption{The variation of $\Delta \chi^2_{\rm NP}= \chi^2_{\rm SM}- \chi^2_{\rm NP}$
  as a function of  $\alpha$  for the non-minimal scenario I (left),
  and as a function of $\beta$ for the non-minimal scenario II (right).}
\label{fig:alpha}
       \end{center}
\end{figure*}

Now we consider the non-minimal scenarios as defined in eqs.~\eqref{eq:non-minimal-eft} and \eqref{eq:non-minimal-eft2}.
The non-minimal scenario I allows additional left-handed $Z$-couplings through
the operator $C_{\phi \ell}^{(1)}$.
On the other hand, in the non-minimal scenario II, right-handed $Z$-boson
couplings are invoked by operator $C_{\phi e}$.
Since neither of these operators contributes to $W$ couplings,
the strong correlation between $W$ and $Z$ couplings,
present in the minimal scenario dominated by $C_{\phi \ell}^{(3)}$, is broken.    
The parameters $\alpha$ and $\beta$ from eqs.~\eqref{eq:non-minimal-eft} and
\eqref{eq:non-minimal-eft2}, respectively, are free parameters in the
context of the EFT, however in specific NP models, they may have fixed
values. As a result, specific NP models may not break the correlation
between $W$ and $Z$ couplings, but just change it.
If we can find the optimal values of these parameters in a model-independent
analysis, it would serve as a guide for the construction of models
for resolving the CAA, at the same time avoiding the tension with the EWP data.

\begin{figure*}[t]
       \begin{center}
\includegraphics[width=0.45\textwidth]{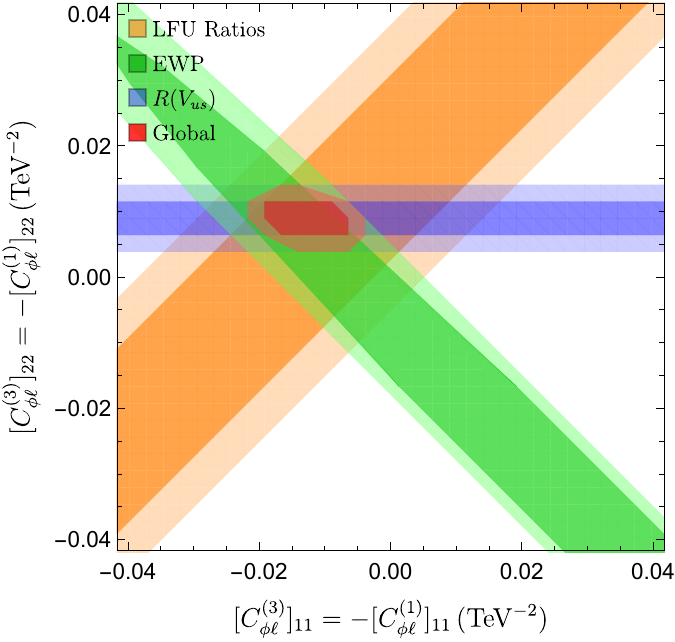}
\hspace{0.65cm}
\includegraphics[width=0.45\textwidth]{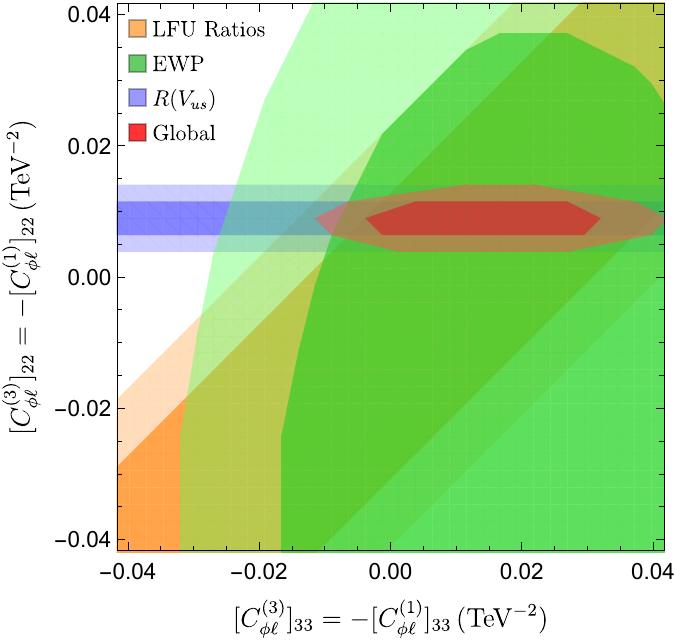} 
\\
\vspace{0.5cm}

\includegraphics[width=0.47\textwidth]{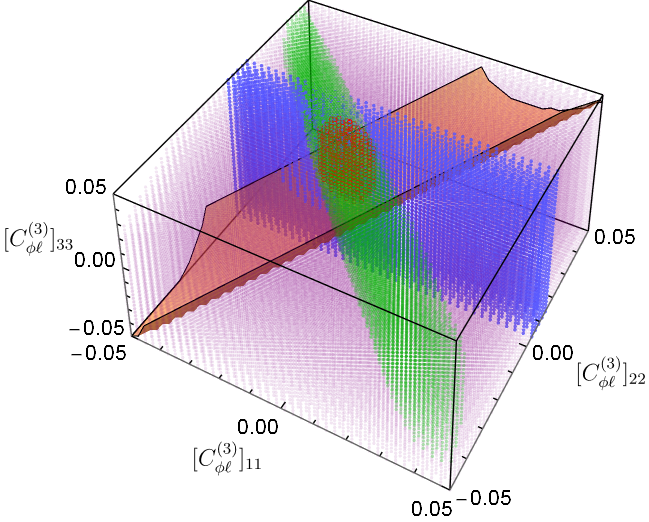}
\hspace{0.5cm}
\includegraphics[width=0.45\textwidth]{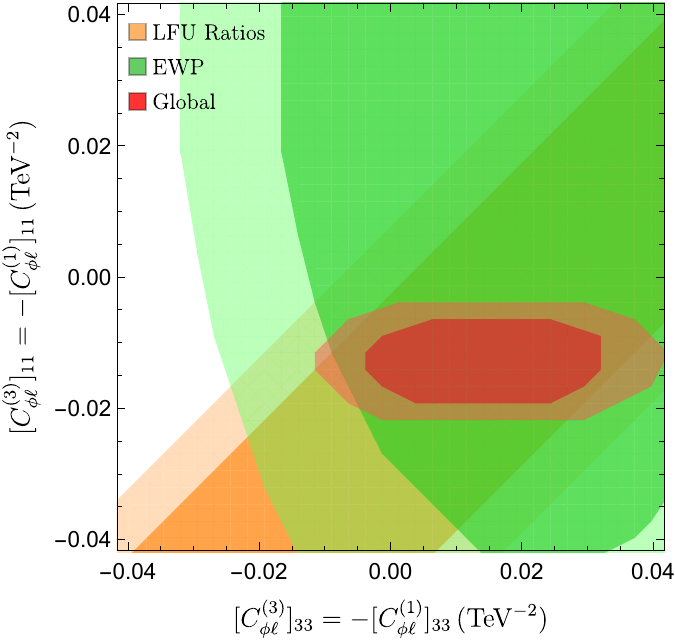}
	       \caption{Same as in Fig. \ref{fig:mi}, except for $\alpha = -1$, i.e. $[C_{\phi \ell }^{(1)}]_{ii}   = -\,  [C_{\phi \ell }^{(3)}]_{ii}$.}
\label{fig:nonminimal}
       \end{center}
\end{figure*}

In order to find out the optimal values of the parameters $\alpha$ and $\beta$,
we study the behavior of $\Delta\chi^2_{\rm NP} \equiv \chi^2_{\rm SM} - \chi^2_{\rm NP}$
as a function of these two parameters. Here $\chi^2_{\rm NP}$ is the minimum
value of $\chi^2$ in the presence of NP in a particular scenario.
The results are shown in Fig.~\ref{fig:alpha}. 

The left panel of Fig.~\ref{fig:alpha} shows that in the non-minimal scenario I
($\alpha \neq 0, \beta=0$), the best global fit is obtained for
 $\alpha \approx -1$, with  $\Delta \chi^2_{\rm NP} \approx 15.0$.
 Thus,  for models with $\alpha \approx -1$, the values of $\chi^2$ will decrease
  by $\approx 4.0$ over the minimal scenario with only $O_{\phi\ell}^{(3)}$.
We illustrate this by showing the allowed regions in the non-minimal
scenario I with $\alpha = -1$, in Fig.~\ref{fig:nonminimal}.
Clearly, all the measurements are consistent with each other, and
with the global best fit, within $1\sigma$.
The favored region has 
$[C_{\phi \ell}^{(1)}]_{22} = -  [C_{\phi \ell}^{(3)}]_{22} < 0$.

From Fig.~\ref{fig:nonminimal}, it may be observed that, the EWP constraints that played a large role in restricting the parameters in the minimal scenario in Fig.~ \ref{fig:mi}  become very weak with $\alpha=-1$, especially in the $\tau-e$ and $\tau-\mu$ planes. As a result, the internal tension between $V_{us}$ observables and the EWP observables is alleviated. This may be attributed to the fact that the NP contribution to $g^\ell_L(Z)$ vanishes for  $\alpha=-1$, so that the impact of NP on the EWP observables reduces. 

The right panel of Fig.~\ref{fig:alpha} shows that in the non-minimal scenario II
($\alpha = 0, \beta \neq 0$), the best global fit is obtained for
 $\beta \approx 0.5$, with $\Delta \chi^2_{\rm NP} \approx 12.1$. Thus, in this class of scenarios, the improvement over the minimal scenario is only marginal. The operator $O_{\phi e}$ cannot decrease the tension between the $V_{us}$ and EWP observables.  

The typical models providing NP corresponding to the non-minimal scenario I
are vector-like lepton (VLL) models, which we will study in the next section.
In order to realize the non-minimal scenario II, one would have to create
models with more particle species.

\section{Minimal vector-like lepton models}
\label{sec:miminal-ext-sm}
\label{subsec:vll-formalism}

In this section, we explore the vector-like lepton (VLL) models, which induce
new tree-level contributions to the $W$ and $Z$ couplings.
These serve as concrete realizations of the non-minimal scenario I
models, which yield specific relations among the non-universal
leptonic couplings of $W$ and $Z$. 
By minimal VLL models, we mean those models that have only a single additional
species of particles in addition to the SM. This species is a vector-like
lepton that couples to the Higgs doublet and the left-handed lepton doublet,
and could be a  singlet or triplet under $SU(2)_L$.
There are four such possible species, whose quantum numbers are given in
Table~\ref{table-vll}. We refer to the models with the name of the
corresponding species, for example the model with an additional species $N$ is referred to as 
Model $N$, etc.

\begin{table*}[thb]
  \begin{center}
\begin{tabular}{cccc}
\toprule
\quad\quad  Vector-like leptons \quad\quad  & \quad \quad\ $SU(3)_C$ \quad\quad &\quad\quad $SU(2)_L$ \quad\quad  & \quad\quad $U(1)_{Y}$ \quad \quad  \\
 \hline
 \quad\quad $N$ \quad\quad  &  \quad\quad 1 \quad\quad & \quad \quad 1 \quad \quad & \quad \quad 0 \quad \quad \\

 \quad\quad $E$ \quad\quad  &  \quad\quad 1 \quad\quad & \quad \quad 1 \quad \quad & \quad \quad -1 \quad \quad \\

 \quad\quad $\Sigma$ \quad\quad  &  \quad\quad 1 \quad\quad & \quad \quad 3 \quad \quad & \quad \quad 0 \quad \quad \\

 \quad\quad $X$ \quad\quad  &  \quad\quad 1 \quad\quad & \quad \quad 3 \quad \quad & \quad \quad -1 \quad \quad \\
 \hline
    \end{tabular}
\caption{The quantum numbers (or the singlet/ triplet nature) of the four VLLs under the SM gauge group.
  Note that in the literature, the notations $\Sigma_0$ and $\Sigma_1$ are sometimes
  used for $\Sigma$ and $X$, respectively \cite{Crivellin:2020ebi}. 
} 
 \label{table-vll}
  \end{center}
\end{table*}

These VLLs can couple to SM Higgs and leptons via the interactions given by
\cite{delAguila:2008pw}
\begin{eqnarray}
\mathcal{L}_{N} & = & (y_N)_i \, \bar N_R \,  \tilde\phi^\dagger \, \ell_{Li}\,,  \\ 
\mathcal{L}_{E} & = & (y_E)_i \,  \bar E_R \,  \phi^\dagger \,  \ell_{Li}\,,  \\ 
\mathcal{L}_{\Sigma} & = & \frac{1}{2}
(y_{\Sigma})_i \, \bar \Sigma_{R}^a \, \tilde \phi^\dagger \, \tau^a \, \ell_{Li}\,,  \\ 
\mathcal{L}_{X} & = & \frac{1}{2} (y_{X})_i \, \bar X_{R}^a  \, \phi^\dagger \, \tau^a
\, \ell_{L_i  }.
\end{eqnarray}
The $\Sigma$ and $X$ leptons, being charged under $ SU(2)_L$, also couple
to the gauge bosons, however that will not affect our analysis.

The couplings of VLLs to the three generations of fermions need not be universal.
If the VLLs are heavy they can be integrated out, leading to the EFT operators like those
discussed in previous sections. This would give rise to effective NP leptonic
couplings of $W$ and $Z$, which would be non-universal.
The couplings in these models may be related to the WCs of the
dimension-6 SMEFT operators as
\begin{equation}
  \label{vll:alpha0}
	[C_{\phi \ell}^{(1)}]_{ij} = \alpha_I \,[C_{\phi \ell}^{(3)}]_{ij}, \quad 
	[C_{\phi \ell}^{(3)}]_{ij} =  {N_I}~   \frac{(y_I)_i^* ~ (y_I)_j}{M_{I}^2}\,,
\end{equation}
where $I$ refers to the relevant VLL species from $\{N, E, \Sigma, X \}$, and
\begin{eqnarray}
  \alpha_N = -1, & \,\,\, \alpha_E = 1, & \,\,\, \alpha_\Sigma = 3, \quad
  \alpha_X = -3 \; ,
  \label{vll-alpha} \\
  N_N = -1/4, & \,\,\,  N_E = -1/4, & \,\,\, N_\Sigma = 1/16, \,\,\,
  N_X = 1/16 \; .
  \label{vll-N}
  \end{eqnarray}
  
 In order to explain $R(V_{us})$, the sign of $[C_{\phi \ell}^{(3)}]_{22}$ needs to be positive. Therefore, even though $\alpha=-1$ for the model $N$, it cannot account for the $V_{us}$ anomaly as $[C_{\phi \ell}^{(3)}]_{22}$ would always be negative, as can be seen from eqs. (\ref{vll:alpha0}) and (\ref{vll-N}). Similarly, model $E$ is of no use for resolving the $V_{us}$ anomaly since it also gives $[C_{\phi \ell}^{(3)}]_{22}<0$. We will therefore only focus on models $X$ and $\Sigma$ in the rest of this paper. 

 Note that the model $\Sigma$ (like model $N$) would also give rise to neutrino masses through the dimension-5 Weinberg operator, and the neutrino masses generated would be too large if we want to have new couplings large enough to account for $R(V_{us})$. The model $\Sigma$  would thus not be a viable solution in its minimal form. However, the neutrino mass problem may be addressed by additional mechanisms \cite{Endo:2020tkb}, so we keep model $\Sigma$  in our further discussions.

\subsection{Validating the model-independent conclusions}

Based on the model-independent results obtained in Sec.~\ref{sec:non-minimal},
the model $X$  with $\alpha_X =-3$ comes close to the optimal value of $\alpha=-1$. This model is therefore
expected to give a much better fit to the data as compared to the other models.
On the other hand, the model $\Sigma$ with $\alpha=+3$, which
is far from the optimal $\alpha$ value, is expected to provide only a marginal improvement over the SM. 
We shall now check if this indeed is the case.

Here, the main difference from the model-independent analysis is that
one needs to take into account the additional
constraints from the LFV decays at the tree level.
In particular, simultaneous presence of $(y_I)_e$ and $(y_I)_\mu$ couplings is
highly constrained by the bounds on the $\mu \to eee$ decay rate.
Therefore, in the following we focus on the 
two cases: $(y_I)_e=0$ and $(y_I)_\mu=0$.
The remaining two non-zero couplings in each case will be related to the WCs, as given in Eq.~\eqref{vll:alpha0}.

Till date, no signals of exotic VLLs have been observed. Since  the VLLs
are pair-produced by the $s$-channel electroweak vector boson diagrams, their
production cross sections are expected to be quite small. 
Therefore, the current bounds on VLLs masses are well below the TeV scale, {\it i.e.,}
$\sim 100$ GeV \cite{Achard:2001qw,Aad:2019kiz,Sirunyan:2019ofn}.
Using a data sample corresponding to an integrated luminosity of 77.4 fb$^{-1}$
of $pp$ collisions at $\sqrt{s}$ = 13 TeV, the CMS collaboration has ruled out
a VLL doublet coupling to the third generation SM leptons in the mass range of
120-790 GeV at 95\% C.L. \cite{Sirunyan:2019ofn}. Obtaining bounds on the masses of
$SU(2)_L$-singlet charged VLLs is an extremely challenging task due to a much
smaller production cross-section and unfavorable branching ratios.
For $SU(2)_L$-triplet VLLs, the future $pp$ colliders such as the 27 TeV high-energy LHC
with 15 ab$^{-1}$ integrated luminosity has the discovery potential
up to 1.7 TeV, whereas a 100 TeV collider with 30 ab$^{-1}$ integrated
  luminosity could make a discovery for masses up to 4 TeV
\cite{Bhattiprolu:2019vdu}. The situation is expected to remain grim for
singlet VLLs. Hence, in the present analysis, we set the mass of VLL particles
to be 1 TeV. The scaling to smaller mass values,  and hence to smaller Yukawa couplings $y_I$, can be obtained through Eq. \eqref{vll:alpha0}.

\begin{figure*}[hbt!]
      \begin{center}
\includegraphics[width=0.45\textwidth]{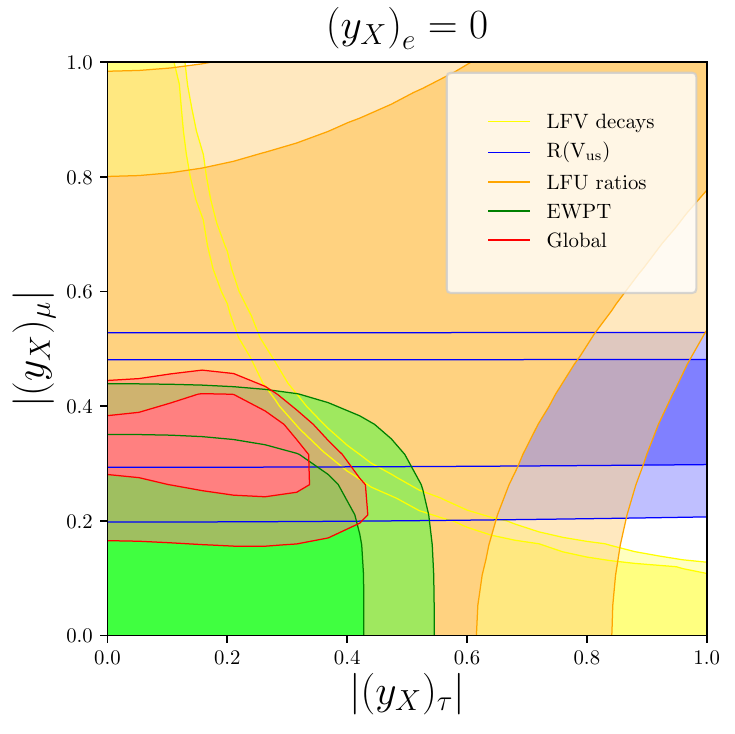}
\hspace{0.5cm}
\includegraphics[width=0.45\textwidth]{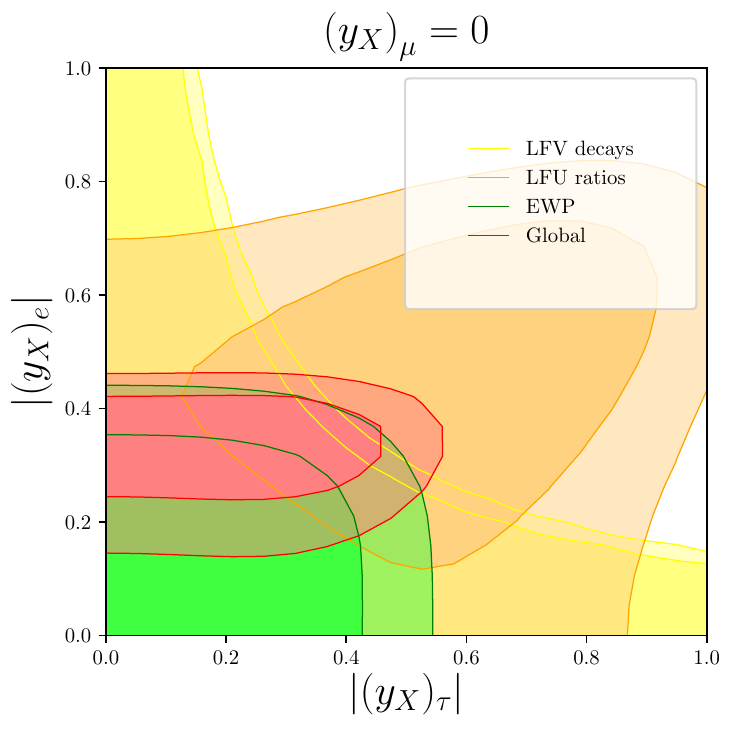}
\caption{Favored regions for Yukawa couplings in the $X$ model.
  Left: constraints in the $\left(|({y_X})_{\mu}|,\,|({y_X})_{\tau}|\right)$ plane, with
  $(y_X)_e =0$. 
  Right: constraints in the $\left(|({y_X})_{e}|,\,|({y_X})_{\tau}|\right)$ plane, with
  $(y_X)_\mu =0$. 
  The dark and light regions correspond to $68\%$ C. L. and $95\%$ C. L,
  respectively. We have taken $M_{X} =1$ TeV. Note that in the right panel, the $R(V_{us})$ measurement plays no role.}
\label{fig:X-model}
       \end{center}
\end{figure*}

The fit results for the $X$ model are shown in Fig.~\ref{fig:X-model}. 
We find that in the $(y_X)_e=0$ case, the best fit clearly favors
non-zero values of $|(y_X)_\mu|$ as well as $|(y_X)_\tau|$.  
The fit also improves significantly over the SM: we have 
$\Delta \chi^2_{\rm NP} \approx 10.0$. 
On the other hand, in the case $(y_X)_\mu=0$, the best fit is very close
to the SM, and the improvement due to NP is only marginal:
$\Delta \chi^2_{\rm NP} \approx 0.5$.
The case with $(y_X)_e=0$ is thus the only case useful for resolving CAA,
and it indeed allows the consistency of all the measurement sets to
within $1\sigma$. A non-zero value of $(y_X)_\mu$ is thus strongly indicated. 
Note that this scenario (model $X$) does not reach the maximum improvement possible with $\alpha=-3$ models as indicated in Fig.~\ref{fig:alpha}, due to the additional constraints coming from the model. However, it does reach close to the model-independent prediction.

\begin{figure*}[hbt!]
      \begin{center}
\includegraphics[width=0.45\textwidth]{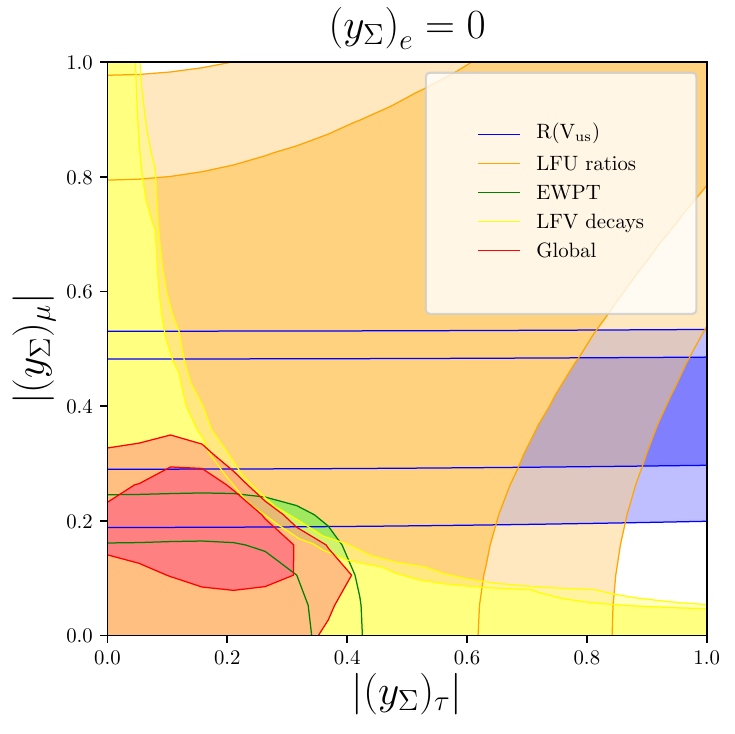}
\hspace{0.5cm}
\includegraphics[width=0.45\textwidth]{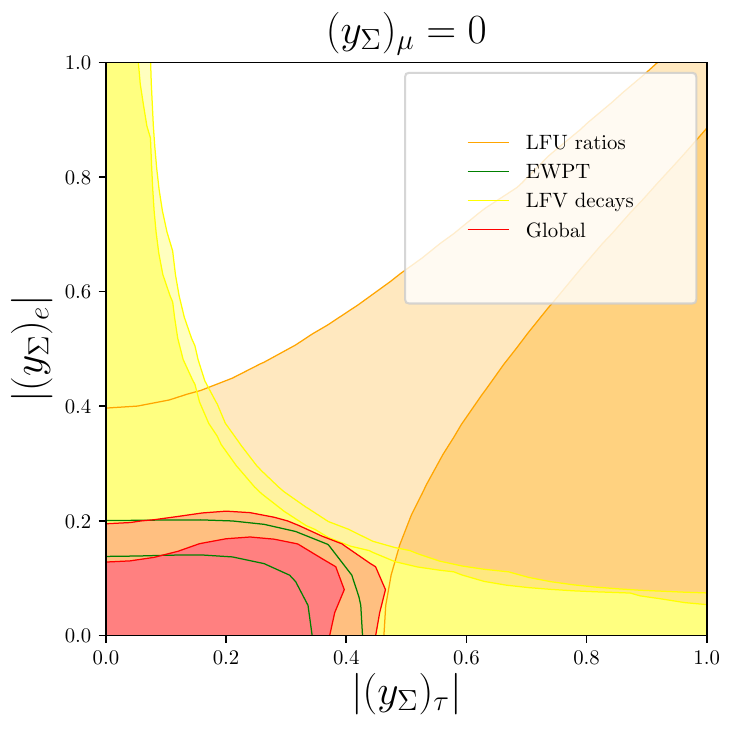}
       \caption{Same as Fig.~\ref{fig:X-model}, but for the $\Sigma$ model.}
\label{fig:Sig-model}
       \end{center}
\end{figure*}

The fit results for the $\Sigma$ model in Fig.~\ref{fig:Sig-model}, on the other hand, are not observed to decrease the 
 tension between the preferred parameter space of EWP data and
$R(V_{us})$ measurement. The 95\%-favored regions
of these two sets of measurements barely overlap.
Thus, as expected from our EFT analysis where the maximum value of $\Delta \chi^2_{\rm NP}$ is 3 for $\alpha = +3$,  this model fails to explain the
data well. This is also reflected in the low values of
$\Delta \chi^2_{\rm NP} =3.3$ and $0.4$, for the case $(y_\Sigma)_e = 0$ and
$(y_\Sigma)_\mu = 0$, respectively. 

We thus find that our model-independent results hold for specific VLL
models, in spite of extra constraints from the
non-observation of LFV decays.

\section{New $W/Z$ couplings and $h \to \ell \ell$ decays}
\label{sec:higgs}

The dimension-6 operators $O_{\phi\ell}^{(1)}$, $O_{\phi\ell}^{(3)}$,
and $O_{\phi e}$, as given in eqs.  \eqref{eq:O1}, \eqref{eq:O3} and
\eqref{eq:Ophie} respectively, involve the coupling of the Higgs doublet
to leptons. However all of these are vector / axial-vector couplings,
and hence are distinct from the Yukawa couplings that give masses to
leptons after EWSB. On the other hand, another dimension-6 operator
composed of the same fields, 
\begin{equation}
  [O_{\phi\ell e}]_{ij} \equiv (\phi^\dagger \phi)(\bar\ell \phi e) \; ,
\end{equation}
modifies the effective strength of Yukawa interactions.\footnote{
  Note that the operator $O_{\phi \ell e}$ is generally denoted in the literature
  as $O_{e \phi}$. We prefer the notation $O_{\phi \ell e}$ as it clearly indicates the
  fields involved, and avoids the possibility of confusion with $O_{\phi e}$.} In the SMEFT framework, this operator has to be present along with the above three, unless some extra symmetry is imposed. In the presence of this operator,
the effective Yukawa interaction becomes
\begin{equation}
  \mathcal{L}_{Y} = -y_{ij} \bar{\ell}_i \,\phi \, e_j +
          [C_{\phi\ell e}]_{ij}\,  [O_{\phi\ell e}]_{ij} 
            + {\rm H.c.} \; ,
\end{equation} 
where $[C_{\phi\ell e}]_{ij}$ are the corresponding WCs.
Note that the leptonic fields here have been written in the flavor basis.

Clearly, the presence of this operator would spoil the relationship between the
mass of a charged lepton and the strength of its coupling with the Higgs boson in SM.
In the basis of mass eigenstates of charged leptons,
\begin{equation}
\delta_{ij} m_{\ell_j} = y_{ij} \dfrac{v}{\sqrt{2}} +
[C_{\phi\ell e}]_{ij} \dfrac{v^3}{2\sqrt{2} \Lambda^2} \; .
\end{equation}
In the SM, we would have the relation
$y_{ij} = \delta_{ij} \cdot \sqrt{2} m_{\ell_j} /v $, which
would imply that the decay width of $h \to \ell^+ \ell^-$ would be
proportional to the square of
$y_\ell \equiv \sqrt{2} m_\ell/v = m_{\ell} \sqrt{G_F^{\cal L}}$.
This, indeed, is one of the precision tests of the SM in the Higgs sector.

Though the operators $O_{\phi\ell}^{(1)}$, $O_{\phi\ell}^{(3)}$,
and $O_{\phi e}$ do not affect the effective Yukawa coupling directly,
$O_{\phi\ell}^{(3)}$ plays a role in spoiling the mass-to-Higgs-coupling
relationship of the charged leptons by modifying the measured
value of $G_F$, and hence the inferred value of $y_{\ell,{\rm SM}} \equiv
m_\ell \sqrt{G_F}$ in terms of which the SM predictions are calculated.
Indeed, even in the absence of $[C_{\phi\ell e}]_{ij}$, one gets
\begin{equation}
  y_{\ell}  
  = m_\ell \sqrt{G_F^{\cal L}}
  = \frac{m_\ell \sqrt{G_F}}
    {\sqrt{1 + \varepsilon_{ee} + \varepsilon_{\mu \mu}}}
  = y_{\ell,\rm SM}\left[1 - \frac{1}{2}(\varepsilon_{ee} +\varepsilon_{\mu\mu})
    \right] ~,
\end{equation}
where $\varepsilon_{i i} = v^2 [C_{\phi\ell}^{(3)}]_{i i}$, as
defined in Eq. \eqref{eq:eps}.  

\begin{figure*}[h!]
       \begin{center}
\includegraphics[height=0.45\textwidth]{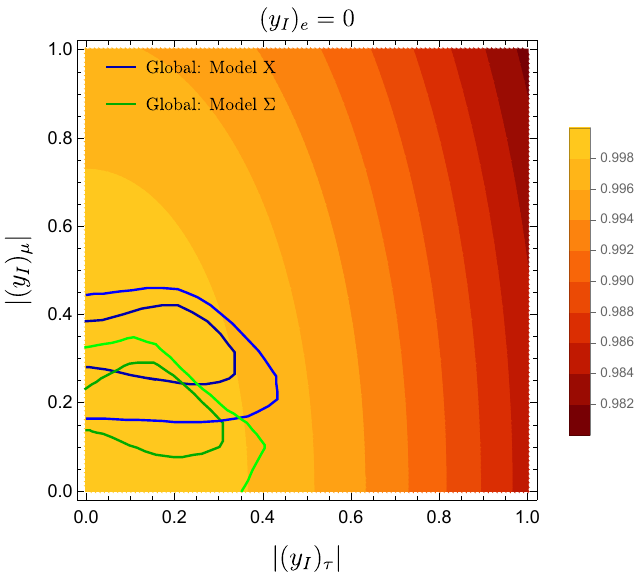}
\hspace{0.5cm}
\includegraphics[height=0.45\textwidth]{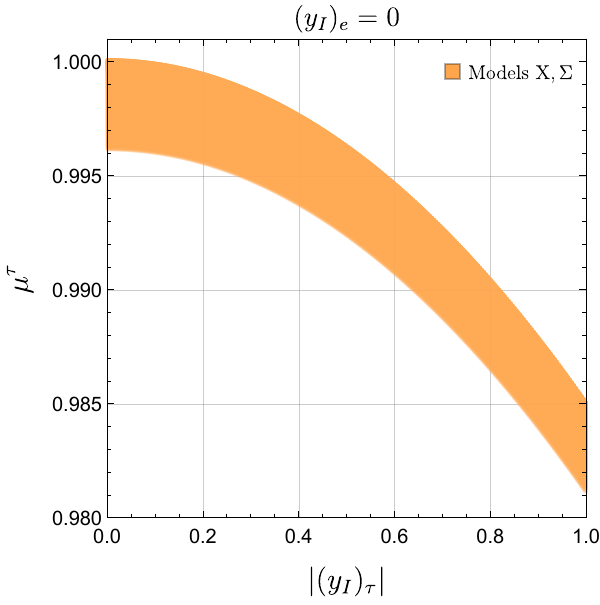}
\caption{Left panel: The signal strength of $h \to \tau \tau$ as a function
  of the Yukawa couplings $|(y_X)_\mu|$ and $|(y_X)_\tau|$. The dark and light blue (green) curves enclose  $68\%$ and $95\%$ C. L favored regions for Yukawa couplings for the $X$ ($\Sigma$) models,
  respectively.
  Right panel: The signal strength of $h \to \tau \tau$ as a function
  of the Yukawa coupling $(y_I)_{\tau}$ for the models $I= X$ and  $\Sigma$.}
\label{fig:higgs}
       \end{center}
\end{figure*}

Combining the above two effects, the signal strength of the Higgs boson
decaying to a pair of leptons is modified by the additional $W$ couplings as
\begin{equation}
  \mu^{i} \equiv \frac{\Gamma(h \to i \,i)}{\Gamma(h \to i\, i)_{\rm SM}} 
  = \Big|1- \frac{1}{2} ( \varepsilon_{\mu \mu} +  \varepsilon_{ee })
  - \frac{1}{y_{\ell,\rm SM}} [C_{\phi\ell e}]_{ii} \Big|^2\; .
  \label{higgs-strength}
\end{equation}
Note that the effect of the $O_{\phi\ell}^{(3)}$ operator is limited to
be very small, since the preferred values of $\varepsilon_{ii}$
are less than a per cent. On the other hand, the $O_{\phi \ell e}$
contribution is enhanced by the inverse of leptonic Yukawa couplings,
and hence can be quite large in the model-independent SMEFT framework.

Let us now explore the minimal VLL models considered earlier, to gauge the
enhancement in the $h \to \ell^+ \ell^-$ decay rate in these models.
It turns out that $[C_{\phi\ell e}]_{ij}$ vanishes in the $N$ model,
while in the $E$, $\Sigma$, and $X$ models, the WCs
$[C_{\phi\ell e}]_{ij}$ are given as  
\begin{align}
  [C_{\phi\ell e}^E]_{ij}
  & = y_{\ell} \left(\frac{(y_E)^*_i (y_E)_j}{2M_E^2}\right)
  = - 2 y_\ell \,  C_{\phi\ell}^{(3)E}\,,
  \label{eq:mu-E} \\ 
 [ C_{\phi\ell e}^{\Sigma}]_{ij}
  & = y_{\ell} \left(\frac{(y_\Sigma)^*_i (y_\Sigma)_j}{8M_\Sigma^2} \right)
 =  2 y_\ell\, C_{\phi\ell}^{(3)\Sigma}\,,
 \label{eq:mu-Sigma}\\ 
  [C_{\phi\ell e}^X]_{ij}
  & = y_{\ell}\left(\frac{(y_X)^*_i (y_X)_j}{8M_X^2}\right)
  =  2 y_\ell \, C_{\phi\ell}^{(3)X} \; .
  \label{eq:mu-X}
\end{align} 
These WCs themselves are thus suppressed by $y_\ell$, consequently
the enhancement shown in Eq. \eqref{higgs-strength} is nullified.

In the left panel of Fig.~\ref{fig:higgs},  we show the signal strength
of the Higgs decay rate to a  pair of $\tau$ leptons, as a function of
the Yukawa couplings $(y_X)_\mu$ and $(y_X)_\tau$ in the model $X$ and $\Sigma$, which are relevant for resolving the $V_{us}$ anomaly..
We take the case $(y_X)_e = 0$, and allow $|(y_X)_\mu|$ and $|(y_X)_\tau|$
to vary in the range $[0-1]$. 
It is observed that the dependence of the signal strength on $(y_X)_\mu$
is almost negligible, compared to its dependence on $(y_X)_\tau$.
We further study the dependence of the signal strength on the latter 
in the right panel of Fig.~\ref{fig:higgs}. Here the results for model $X$ and model $\Sigma$ are identical, 
since the functional dependences of the signal strength on
$(y_I)_\mu$ and $(y_I)_\tau$ are identical for these two models,
as can be seen from eqs. \eqref{eq:mu-Sigma} and \eqref{eq:mu-X}. 
It is observed that
the value of the signal strength $\mu^{\tau}$ can become less than unity, however the change is less than a few per cent for allowed values of $(y_I)_\mu$ and $(y_I)_\tau$.

The effect on the branching fraction $h \to \tau\tau$ is thus very small in the minimal
VLL models, though it may be probed at future precision machines.
For example, the HL-LHC can measure the $h\tau\tau$ coupling up to
a per cent level precision, whereas the $e^+ e^-$ Higgs factories such
as TLEP can probe it to below a per cent level \cite{Gomez-Ceballos:2013zzn}. 
The more exciting possibility, however, is the discovery of a larger
deviation than a few per cent, which will rule out the minimal VLL
models and point towards a more interesting scenario.

\section{Conclusions}
\label{sec:conc}

The observed discrepancies between determinations of $V_{us}$ from different measurements such as semileptonic kaon decays, $\beta$ decays and $\tau$ decays, may 
be viewed as signatures of LFU violation in the $W$-boson couplings to  leptons. In this work, we explore this Cabibbo angle anomaly (CAA)   within the SMEFT framework by considering  gauge invariant dimension-6 operators, which modify the couplings of leptons to the $W$ and $Z$ bosons at the tree level. One of these gauge invariant operators, $O_{\phi \ell}^{(3)}$, which modifies the $W$ couplings, is essential to resolve this anomaly. We perform a model-independent global fit to the EWP data, measurements of several LFU observables, and various measurements of $V_{us}$, in order to check the consistency among these sets of measurements.  
We show that a tension exists between the parameter space favored by EWP data and the solution of the CAA, if we only have the  operator $O_{\phi \ell}^{(3)}$. This is due to the fact that the operator $O_{\phi \ell}^{(3)}$, needed for resolving the CAA, also induces corrections to the $Z$-couplings, and hence is highly constrained by the EWP data.

We then show that the above tension  can be alleviated by introducing additional sources of gauge-invariant couplings of $Z$ boson to the 
left- and right-handed leptons via the 6-dimensional operators $O_{\phi \ell}^{(1)}$ and $O_{ \phi e}$. We find that the optimal solution which resolves the tension  corresponds to specific relations between the WCs of these operators given by: $C_{\phi \ell}^{(1)} \approx -\, C_{\phi \ell}^{(3)} $, or $C_{\phi e} \approx 0.5\, C_{\phi \ell}^{(3)}$, when
either of the two operators $O_{\phi \ell}^{(1)}$ or $O_{ \phi e}$ is present at a time. 
The condition of  $C_{\phi \ell}^{(1)} \approx -\, C_{\phi \ell}^{(3)} $ yields vanishing NP coupling of $Z$ to the lepton doublet, thus reducing the effect of NP on the EWP observables, and hence is the most successful in alleviating the tension.

In order  to illustrate the implications of our model independent results, we analyze minimal extensions of the SM involving the VLL models. We consider the models with one of the $SU(2)_L$ singlets $N$ and $E$, or one of the $SU(2)_L$ triplets $\Sigma$ and $X$.  Out of these four VLL models, the models $N$ and $E$ cannot resolve the CAA since they lead to an opposite sign for the NP contribution than what is needed. The model $X$, with parameters close to the optimal ones implied by the model-independent analysis,  alleviates the tension between $V_{us}$ and EWP observables, whereas  the model $\Sigma$ is seen to provide only a marginal improvement over the SM as it is far from the optimal scenario.

 Finally, we study the impact of a related new 6-dimensional operator $O_{\phi \ell e}$  on the signal strength of the Higgs boson decay to a pair of leptons. This operator is mandatory in the SMEFT framework in the absence of any extra symmetry. We find that this operator can affect the signal substantially in a general NP scenario. However, for the favored parameter space of the minimal VLL models, these signal strengths can be modified only to less than 1\%. This may be  accessible at the future Higgs factories for the $h \to \tau \tau$ decay mode.  However, an exciting possibility would be to find more than a few per cent deviation from the predicted SM decay rate, which will indicate the presence of non-minimal VLL models that give rise to a large $O_{\phi \ell e}$.
 
 \section{Acknowledgments}
JK acknowledges financial support from NSERC of Canada. AKA and SG would like to thank Kazuki Sakurai for useful discussions. AD and SG acknowledge support of the Department of Atomic Energy (DAE), Government of India, under Project Identification No. RTI4002.


\end{document}